\documentclass[12pt]{article}
\usepackage[active]{srcltx}
\usepackage{epsfig}
\usepackage{graphicx}
\usepackage{cite}
\newcommand{\eref}[1]{(\ref{#1})}

\newcommand{\na}{\mbox{\boldmath $\nabla$}}

\newcommand{\E}{\mbox{\boldmath $E$}}

\newcommand{\Hb}{\mbox{\boldmath $H$}}
\newcommand{\av}{\mbox{\boldmath $a$}}
\newcommand{\Av}{\mbox{\boldmath $A$}}

\newcommand{\cE}{\mbox{\boldmath$\cal E$}}
\newcommand{\cH}{\mbox{\boldmath$\cal H$}}
\newcommand{\Rho}{\mbox{\boldmath $\hat{\cal R}$}}
\newcommand{\Ta}{\mbox{\boldmath $\hat{\cal T}$}}

\renewcommand{\t}{\mbox{\boldmath $t$}}
\newcommand{\D}{\mbox{\boldmath $D$}}

\newcommand{\eps}{\mbox{\boldmath $\varepsilon$}}
\newcommand{\ka}{\mbox{\boldmath $\kappa$}}
\newcommand{\ps}{\mbox{\boldmath $\psi$}}

\newcommand{\e}{\mbox{\boldmath $e$}}

\newcommand{\n}{\mbox{\boldmath $n$}}

\newcommand{\rr}{\mbox{\boldmath $r$}}

\newcommand{\J}{\mbox{\boldmath $J$}}

\newcommand{\kk}{\mbox{\boldmath $k$}}

\newcommand{\lv}{\mbox{\boldmath$l$}}
\newcommand{\hI}{\mbox{\boldmath$\hat{I}$}}

\newcommand{\hE}{\mbox{\boldmath $\hat{E}$}}
\newcommand{\hT}{\mbox{\boldmath $\hat{T}$}}
\newcommand{\hR}{\mbox{\boldmath $\hat{R}$}}

\newcommand\fr{\displaystyle\frac}

\newcommand\ola{\overleftarrow}
\newcommand\ora{\overrightarrow}
\newcommand{\htts}{\mbox{\boldmath$\hat{t}\kern1pt$}}

\newcommand\lt{\left}
\newcommand\rt{\right}

\newcommand{\skk}{\mbox{\scriptsize\boldmath$k$}}
\newcommand{\srr}{\mbox{\scriptsize\boldmath$r$}}

\begin{document}
\begin{center}
Ignatovich F.V.$^1$, Ignatovich V.K.$^{2*}$,

$^1$ Lumetrics inc, Rochester, N.Y., USA

$^2$FLNP JINR, Dubna, Russia
\bigskip

{\Large\bf Reflection of light from an anisotropic medium}
\end{center}

\begin{abstract}
We present here a general approach to treat reflection and
refraction of light of arbitrary polarization from single axis
anisotropic plates. We show that reflection from interface inside
the anisotropic medium is accompanied by beam splitting and can
create surface waves.
\end{abstract}

\section{Introduction}

Theory of anisotropic media in optics alienates by its complexity,
phenomenology and not transparent
physics~\cite{fed,kuz,kuz1,dit,land}. The most recent paper
related to this field is published in Am.J.Phys.-\cite{kunz} only
in 1977. It is understandable that no one likes to discuss this
topic. In the science of almost two hundred years old it seems
hopeless to propose something new which can be attractive to
students. Nevertheless, in this paper we will do just that. We
will present an approach similar to that, used for description of
elastic waves in anisotropic media~\cite{igl}

   The single axis anisotropic nonmagnetic media in electromagnetism
can be described by a matrix of dielectric permittivity $\eps$
with matrix elements~\cite{fed}
\begin{equation}\label{1}
\eps_{ij}=\epsilon_0\delta_{ij}+\epsilon'a_ia_j,
\end{equation}
where $\epsilon_0$ is isotropical part, and anisotropy is
characterized by the unit vector $\av$ with components $a_i$ and
by anisotropy parameter $\epsilon'$. With this matrix we will
first consider propagation of electromagnetic waves in a
homogeneous anisotropic medium, then refraction at the interface
between anisotropic and isotropic media, and after that discuss
reflection and transmission of an anisotropic plain plate of
finite thickness. Time to time along the text we make digression
to isotropic media, because on one side it helps to check
correctness of our formulas, and on the other side it is useful,
because isotropic media also represent some difficulties.

Of course, some our formulas look huge, but we don't care about
it. We do not even derive some formulas and leave it to those who
will need it. In fact, it is sufficient to know how to derive
them, and if necessary to use a computer.

\section{Waves in an anisotropic medium}

The wave equation for, say, electric field $\E(\rr,t)$, is
obtained from Maxwell equations. In a homogeneous nonmagnetic
($\mu=1$) anisotropic medium it is
\begin{equation}\label{2}
-[\na\times[\na \times \E(\rr,t)]]= \fr{\partial^2}{c^2\partial
t^2}\eps\E(\rr,t),
\end{equation}
Solution of this equation can be accepted in the form of a plain
wave
\begin{equation}\label{3}
\E(\rr,t)=\cE\exp(i\kk\cdot\rr-i\omega t),
\end{equation}
where $\cE$ is a polarization vector, which can be of not unit
length.

After substitution of \eref{3} into \eref{2} we obtain
\begin{equation}\label{4}
k^2\cE-\kk(\kk\cdot\cE)=k_0^2\eps\cE,
\end{equation}
where $k_0=\omega/c$, and $\eps\cE$, according to \eref{1} is
\begin{equation}\label{a4}
\eps\cE=\epsilon_0\cE+\epsilon'\av(\av\cdot\cE).
\end{equation}

In isotropic media polarization vector $\cE$ can have arbitrary
direction perpendicular to the wave vector $\kk$. In anisotropic
media it is not so.

Besides \eref{a4} the field $\E(\rr,t)$ should also satisfy
Maxwell equation
\begin{equation}\label{5}
\na\cdot\eps\E(\rr,t) = 0,
\end{equation}
from which, after substitution of \eref{3}, it follows a
limitation on possible choices of directions for $\cE$:
\begin{equation}\label{6}
\epsilon_0(\kk\cdot\cE)+\epsilon'(\kk\cdot\av)(\av\cdot\cE)=0.
\end{equation}

If $\kk$ is not parallel to $\av$, we have three independent
vectors $\av$, $\ka=\kk/k$ and $\e_1=[\av\times\ka]$, which can
serve a basis of a coordinate system. The basis is not
orthonormal, nevertheless the vector $\cE$ can be represented in
this basis as
\begin{equation}\label{8}
\cE=\alpha\av+\beta\ka+\gamma\e_1
\end{equation}
with some coordinates $\alpha$, $\beta$ and $\gamma$, which are
not arbitrary, as will be now shown.

Substitution of \eref{8} into \eref{6} gives
\begin{equation}\label{9}
\epsilon_0[k\beta+(\kk\cdot\av)\alpha]+\epsilon'(\kk\cdot\av)[\alpha+\beta(\ka\cdot\av)]=0,
\end{equation}
 from which it follows that
\begin{equation}\label{10}
\beta=-\fr{(\ka\cdot\av)(1+\eta)} {1+\eta(\ka\cdot\av)^2}\alpha,
\end{equation}
where $\eta=\epsilon'/\epsilon_0$. Substitution of \eref{10} into
\eref{8} gives
\begin{equation}\label{11}
\cE=\alpha\lt(\av-\ka\fr{(\ka\cdot\av)(1+\eta)}
{1+\eta(\ka\cdot\av)^2}\rt)+\gamma\e_1=\alpha\e_2+\gamma\e_1,
\end{equation}
which shows that in fact we have only two independent vectors for
expansion of $\cE$: vector $\e_1=[\av\times\ka]$ orthogonal to the
plane of $\av,\kk$, and the vector
\begin{equation}\label{12}
\e_2=\av-\ka\fr{(\ka\cdot\av)(1+\eta)} {1+\eta(\ka\cdot\av)^2}.
\end{equation}

It is worth to note that $\e_2$ is not an eigen vector of matrix
$\eps$, but the matrix $\eps$ transforms $\e_2$ into a vector
orthogonal to $\kk$
\begin{equation}\label{1a2}
\eps\e_2=\epsilon_1(\theta)[\ka\times[\av\times\ka]],
\end{equation}
where
\begin{equation}\label{1b2}
\epsilon_1(\theta)=\epsilon_0\fr{1+\eta}{1+\eta(\ka\cdot\av)^2}=\epsilon_0\fr{1+\eta}{1+\eta\cos^2\theta},
\end{equation}
and $\theta$ is the angle between vectors $\av$ and $\kk$. It is
seen that, if the wave propagates along $\av$, the dielectric
permittivity becomes $\epsilon_1(0)=\epsilon_0$, and, if the wave
propagates perpendicularly to $\av$, the dielectric permittivity
becomes $\epsilon_1(\pi/2)=\epsilon_0+\epsilon'$.

With account of \eref{1b2} we can represent \eref{12} as
\begin{equation}\label{1c2}
\e_2=\av-\ka(\ka\cdot\av)\epsilon_1(\theta)/\epsilon_0.
\end{equation}
In the limit $\eta\to0$, when the medium becomes isotropic we
obtain
\begin{equation}\label{1c2a}
\lim_{\eta\to0}\e_2=[\ka\times[\av\times\ka]].
\end{equation}

Now we will show that in general a plain wave in anisotropic media
can have polarizations only along $\e_2$ or $\e_1$.

Indeed, substitution of \eref{11} into \eref{4} with account of
\eref{1a2} gives
\begin{equation}\label{a11a}
k^2[\alpha\e_2+\gamma\e_1-\alpha\ka(\ka\cdot\e_2)]=
k_0^2\Big(\alpha\epsilon_1(\theta)[\ka\times[\av\times\ka]]+\gamma\epsilon_0\e_1\Big).
\end{equation}
Multiplying it with $\e_1$ we obtain
\begin{equation}\label{12b}
(k^2-k_0^2\epsilon_0)\gamma\e_1^2=0.
\end{equation}
Therefore, if $\gamma\ne0$, \eref{12b} can be satisfied only, when
\begin{equation}\label{13}
k^2=k_0^2\epsilon_0.
\end{equation}
Multiplying \eref{a11a} with $\e_2$ we obtain
\begin{equation}\label{12c}
\lt(k^2-k_0^2\epsilon_1(\theta)\rt)\alpha[1-(\av\cdot\ka)^2]=0.
\end{equation}
Therefore, if $\alpha\ne0$, and $\av\ne\ka$, \eref{12c} can be satisfied only, when
\begin{equation}\label{12d}
k^2=k_0^2\epsilon_1(\theta),
\end{equation}
where $\epsilon_1(\theta)$ is given in \eref{1b2}. Since the
length of $k$ is different for two polarization vectors, therefore
a single plain wave can exist only with a single polarization
along either $\e_2$ or $\e_1$.

We will call ``transverse'' the mode with polarization
$\cE_1=\e_1$, and ``mixed'' the mode with polarization along
$\cE_2=\e_2$. The mixed mode contains a longitudinal component --
a component along the wave vector $\kk$. We think that such a
nomenclature is better than common names: ``ordinary'' for
$\cE_1=\e_1$, and ``extraordinary'' for $\cE_2=\e_2$, because our
names point to physical peculiarities of these waves.

\subsection{Magnetic fields}

Every electromagnetic wave besides electric contains also magnetic
field. For simplicity we assume that $\mu=1$. From the equation
$\na\cdot\Hb=0$, which is equivalent to $\kk\cdot\Hb=0$, it
follows that the field $\Hb$ is orthogonal to $\kk$. It is also
orthogonal to $\cE$, which follows from the Maxwell equation
\begin{equation}\label{3a}
-[\na \times \E(\rr,t)] = \frac{\partial}{c\partial t}\Hb(\rr,t).
\end{equation}
After substitution of \eref{3} and
\begin{equation}\label{3a2}
\Hb(\rr,t)=\cH\exp(i\kk\cdot\rr-i\omega t),
\end{equation}
with polarization vector $\cH$ of the field $\Hb$, we obtain
\begin{equation}\label{3a1}
\cH=\fr k{k_0}[\ka\times\cE],
\end{equation}
where $k_0=\omega/c$, and $\ka=\kk/k$ is a unit vector along the
wave vector $\kk$. For transverse and mixed modes, respectively,
we therefore obtain
\begin{equation}\label{16a}
\cH_1=\fr
k{k_0}[\ka\times\e_1]=\fr{k}{k_0}[\ka\times[\av\times\ka]],\qquad
\cH_2=\fr{k}{k_0}[\ka\times\e_2]=\fr{k}{k_0}[\ka\times\av],
\end{equation}
and the total plain wave field looks
\begin{equation}\label{16a1}
\Psi(\rr,t)=\psi_j\exp(i\kk_j\cdot\rr-i\omega t),
\end{equation}
where $\psi_j=\cE_j+\cH_j$, and $j$ denotes mode 1 or 2. In
isotropic media we also can choose, say $\cE=[\av\times\ka]$ and
$\cH=[\ka\times[\av\times\ka]]$. However, there $\av$ can have
arbitrary direction, therefore the couple of orthogonal vectors
$\cE$ and $\cH$ can be rotated any angle around the wave vector
$\kk$.

\section{Reflection from an interface with an isotropic medium}

Imagine that our space is split into two half spaces. The part at
$z<0$ is an anisotropic medium, and the part at $z>0$ is vacuum
with $\epsilon_0=1$, $\eta=0$. We have two different wave
equations in these parts, and waves go from reign of one equation
into reign of another one through the interface where they must
obey boundary conditions imposed by Maxwell equations.

Let's look for reflection of the two possible modes incident onto
the interface within the anisotropic medium.

\subsection{Nonspecularity and mode transformation at the interface}

First we note that reflection of mixed mode is not specular.
Indeed, since direction of $\kk$ after reflection changes,
therefore the angle $\theta$ between $\av$ and $\ka$ does also
change, and $k$, according to \eref{12d}, changes too. However the
component $\kk_\|$ parallel to the interface does not change, so
the change of $k$ means the change of the normal component
$k_\bot$, and this leads to nonspecularity of the reflection.

Let's calculate the change of $k_\bot$ for the incident mixed mode
with wave vector $\kk_{2r}$, where the index r means that the mode
2 propagates to the right toward the interface. For a given angle
$\theta$ between $\kk_{2r}$ and $\av$ we can write
\begin{equation}\label{17}
k_{2r\bot}=\sqrt{\fr{\epsilon_0k_0^2(1+\eta)}{1+
\eta\cos^2\theta}-k_\|^2},
\end{equation}
however the value of $k_{2r\bot}$ enters implicitly into
$\cos\theta$, so to find explicit dependence of $k_{2r\bot}$ on
$\av$ it is necessary to solve the equation
\begin{equation}\label{18}
k_\|^2+x^2+\eta(k_\|(\lv\cdot\av)+x(\n\cdot\av))^2=k_0^2\epsilon_0(1+\eta),
\end{equation}
where $x$ denotes $k_{2r\bot}$, $\n$ is a unit vector of normal,
directed toward isotropic medium, and $\lv$ is a unit vector along
$\kk_\|$, which together with $\n$ constitutes the plane of
incidence. Solution of this equation is
\begin{equation}\label{19}
x=\fr{-\eta
k_\|(\n\cdot\av)(\lv\cdot\av)+\sqrt{\epsilon_0k_0^2(1+\eta)(1+\eta(\n\cdot\av)^2)-
k_\|^2(1+\eta(\lv\cdot\av)^2+\eta(\n\cdot\av)^2)}}{1+\eta(\n\cdot\av)^2}.
\end{equation}
The sign chosen before square root provides the correct
asymptotics at $\eta=0$ equal to isotropic value
$\sqrt{\epsilon_0k_0^2- k_\|^2}$.

In general vector $\av$ is representable as
$\av=\alpha\n+\beta\lv+\gamma\t$, where $\t=[\n\lv]$ is a unit
vector perpendicular to the plane of incidence. The normal
component $k_{2r\bot}$ depends only on part of this vector
$\av'=\alpha\n+\beta\lv$, which lies in the incidence plane. If we
denote $\alpha=|\av'|\cos(\theta_a)$,
$\beta=|\av'|\sin(\theta_a)$, where $|\av'|$ is projection of
$\av$ on the incidence plane, and introduce new parameter
$\eta'=\eta|\av'|^2\le\eta$, then formula \eref{19} is simplified
to
\begin{equation}\label{19a}
k_{2r\bot}=\fr{-\eta'
k_\|\sin(2\theta_a)+2\sqrt{\epsilon_0k_0^2(1+\eta)[1+\eta'\cos^2(\theta_a)]-
k_\|^2(1+\eta')}}{2[1+\eta'\cos^2(\theta_a)]}.
\end{equation}

For the reflected mixed mode (mode 2, propagating to the left from
the interface) an equation similar to \eref{18} looks
\begin{equation}\label{20}
k_\|^2+x^2+\eta(k_\|(\lv\cdot\av)-x(\n\cdot\av))^2=k_0^2\epsilon_0(1+\eta),
\end{equation}
where $x=k_{2l\bot}$, and its solution is
\begin{equation}\label{21}
k_{2l\bot}=\fr{\eta'
k_\|\sin(2\theta_a)+2\sqrt{\epsilon_0k_0^2(1+\eta)[1+\eta'\cos^2(\theta_a)]-
k_\|^2(1+\eta')}}{2[1+\eta'\cos^2(\theta_a)]}.
\end{equation}
We see that the difference of the normal components of reflected
and incident waves of mixed modes $k_{2l\bot}-k_{2r\bot}$ is
\begin{equation}\label{22}
k_{2l\bot}-k_{2r\bot}=\fr{\eta'
k_\|\sin(2\theta_a)}{1+\eta'\cos^2(\theta_a)},
\end{equation}
In the following we will present such differences in dimensionless
variables
\begin{equation}\label{a22}
\Delta_{22}\equiv\fr{k_{2l\bot}-k_{2r\bot}}{k_0\sqrt{\epsilon_0}}=\fr{\eta'
q\sin(2\theta_a)+2\sqrt{(1+\eta)[1+\eta'\cos^2(\theta_a)]-
q^2(1+\eta')}}{2[1+\eta'\cos^2(\theta_a)]},
\end{equation}
where $q^2=k_\|^2/k_0^2\epsilon_0$. The reflection angle depends
on orientation of anisotropy vector $\av$ and it can be both
larger than the specular one, when $\theta_a>0$, or smaller, when
$\theta_a<0$.

In the case of transverse incident mode the length $k=|\kk|$ of
the wave vector, according to \eref{13}, does not depend on
orientation of $\av$, therefore this wave is reflected specularly.
\begin{figure}[t!]
{\par\centering\resizebox*{8cm}{!}{\includegraphics{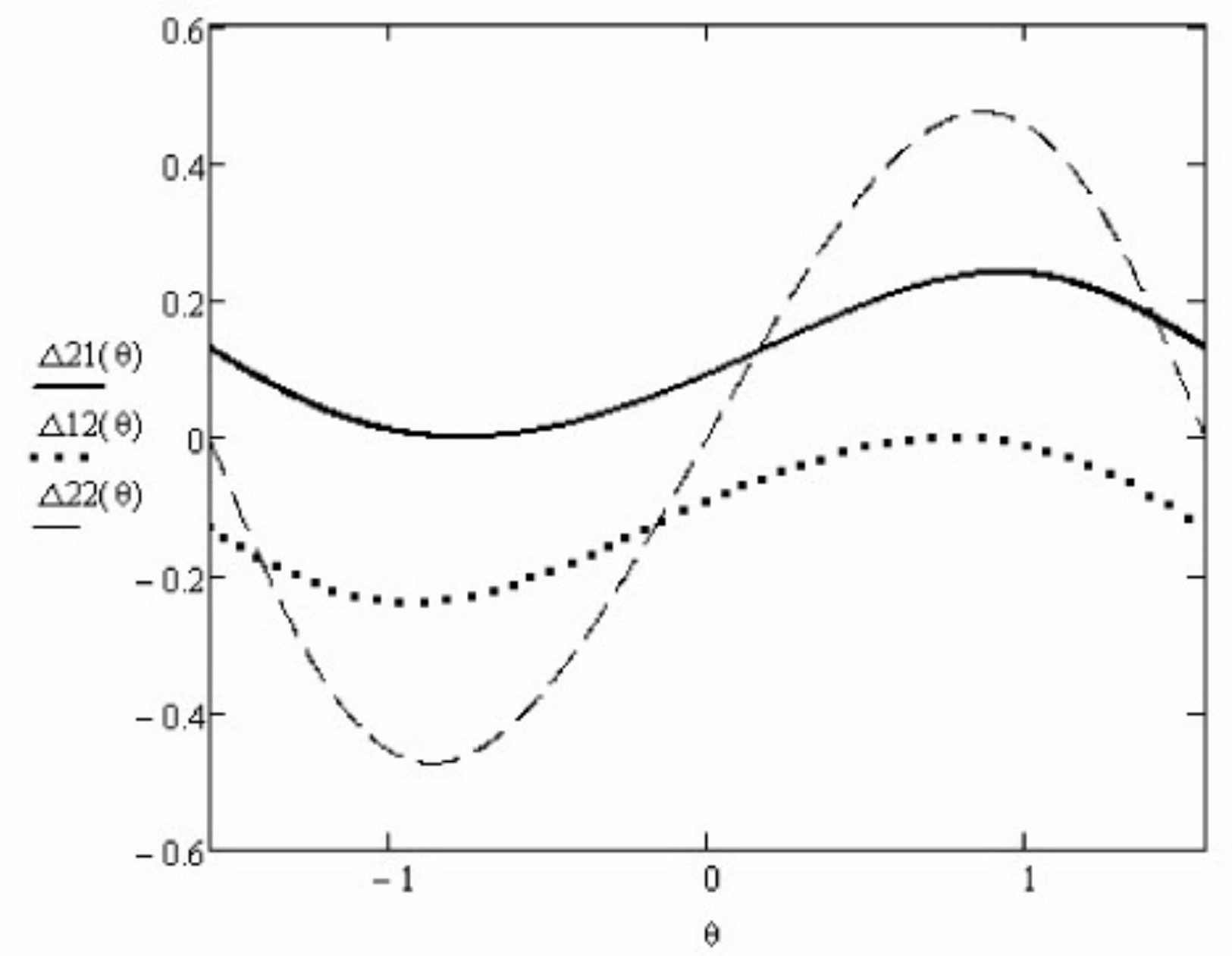}}\par}
\caption{Variation of change $\Delta$ of normal components for
reflected and incident waves in dependence of angle
$\theta=\theta_a$ of anisotropy vector with respect to normal
$\n$. Vector $\av$ is supposed to lie completely in the incidence
plane. The curves $\Delta ij$ represent dimensionless ratio
$\Delta_{ij}(\theta)$ given by \eref{a22}, \eref{22a} and
\eref{22b}. The curves were calculated for $\eta=\eta'=0.4$ and
$q=k_\|/k_0\sqrt{\epsilon_0}=0.7$.} \label{f1}
\end{figure}

Every incident mode after reflection creates another one, without
another mode it is impossible to satisfy the boundary conditions.
Let's look what will be the normal component of the other mode. If
the incident is of mode 2, reflected transverse mode (mode 1
propagating to the left, away from the interface) will have
$k_{1l\bot}=\sqrt{\epsilon_0k_0^2-k_\|^2}$. Therefore according to
\eref{19a} the difference
$\Delta_{12}=(k_{1l\bot}-k_{2r\bot})/k_0\sqrt{\epsilon_0}$ is
\begin{equation}\label{22a}
\Delta_{12}=\sqrt{1-q^2}-\fr{-\eta'
q\sin(2\theta_a)+2\sqrt{(1+\eta)[1+\eta'\cos^2(\theta_a)]-
q^2(1+\eta')}}{2[1+\eta'\cos^2(\theta_a)]}.
\end{equation}
In the opposite case, when the incident mode is transverse one,
the reflected mixed mode will have $k_{2l\bot}$ shown in
\eref{21}. Therefore the difference
$\Delta_{21}=(k_{2l\bot}-k_{1r\bot})/k_0\sqrt{\epsilon_0}$ is
\begin{equation}\label{22b}
\Delta_{21}=\fr{\eta'
q\sin(2\theta_a)+2\sqrt{(1+\eta)[1+\eta'\cos^2(\theta_a)]-
q^2(1+\eta')}}{2[1+\eta'\cos^2(\theta_a)]}-\sqrt{1-q^2}.
\end{equation}
The changes of normal components with variation of $\theta_a$
according to \eref{a22}, \eref{22a} and \eref{22b} for some values
of dimensionless parameters $\eta$ and $q$ and vector $\av$ lying
completely in the incidence plane, are shown in Fig. \ref{f1}.
From this figure it is seen that the strongest deviation of
reflected wave from specular direction is observed for reflection
of mixed to mixed mode.
\begin{figure}[t!]
{\par\centering\resizebox*{6cm}{!}{\includegraphics{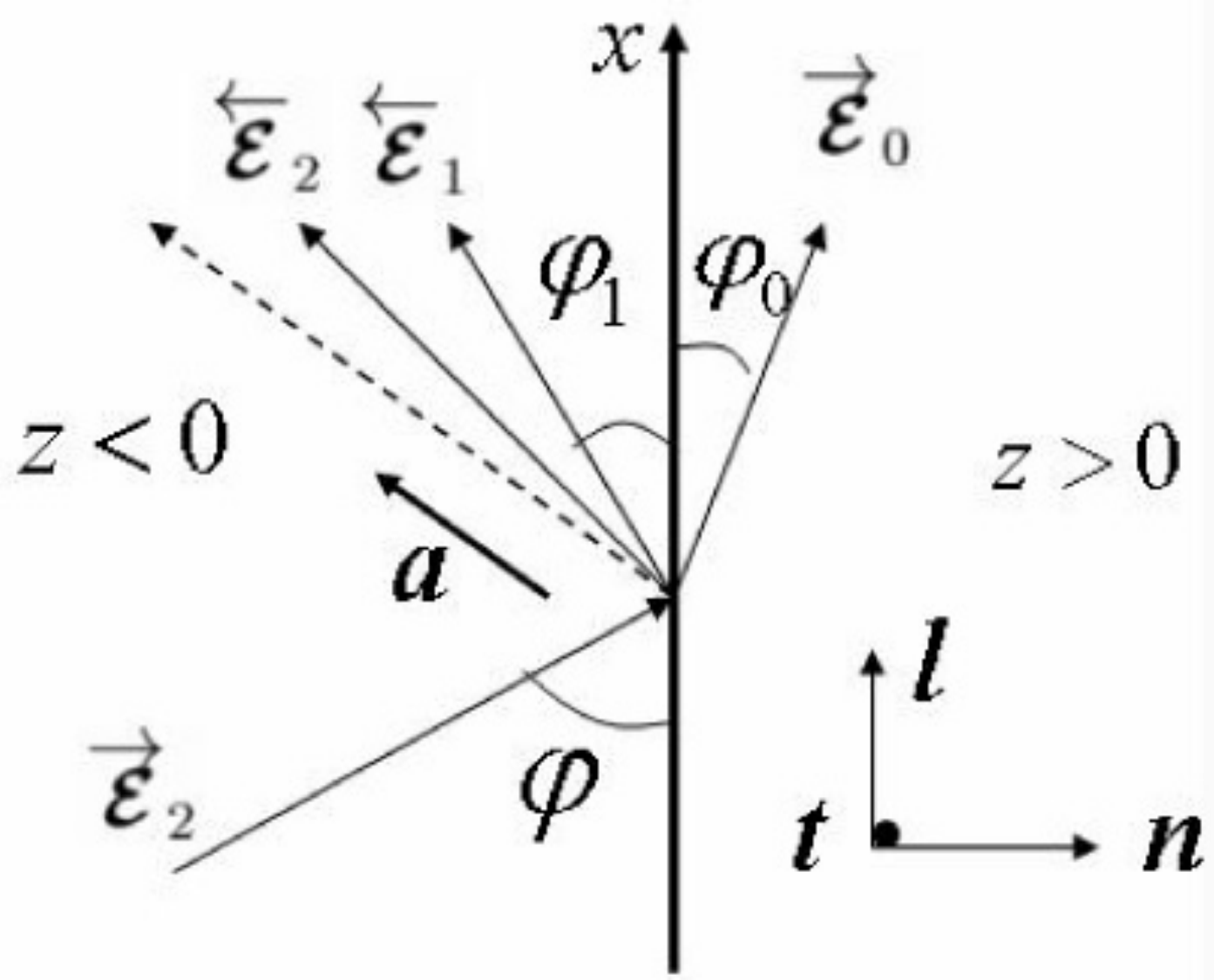}}\par}
\caption{Arrangement of wave vectors of all the modes resulted
when the incident wave is of mode 2, $\protect\ora\cE$, and when
the anisotropy vector $\av$ has the direction as shown here. The
grazing angle of the reflected mode 2, $\protect\ola\cE_2$, is
less than specular one (specular direction is shown by broken
arrow), and the grazing angle $\varphi_1$ of the reflected mode 1,
$\protect\ola\cE_1$, is even lower. The grazing angle $\varphi_0$
of the transmitted wave $\protect\ora\cE_0$ is even lower than
$\varphi_1$. We can imagine than at some critical value
$\varphi=\varphi_{c1}$ the angle $\varphi_0$ becomes zero. It
means that at $\varphi<\varphi_{c1}$ transmitted wave becomes
evanescent and all the incident energy is totally reflected in the
form of two modes. More over, there exists a second critical angle
$\varphi_{c2}$, when $\varphi_1=0$. It means that at
$\varphi<\varphi_{c2}$ the mode $\protect\ola\cE_1$ also becomes
evanescent. In this case all the incident energy is totally
reflected nonspecularly in the form of mode 2. At the same time
the two evanescent waves $\protect\ora\cE_0$ and
$\protect\ola\cE_2$ combine into a surface wave, propagating along
the interface. The arrows over $\cE$ show direction of waves
propagation with respect to the interface. In the figure there is
also shown the basis which is used along the paper. It consists of
unit normal vector $\n$ along normal ($z$-axis), unit tangential
vector $\lv$ ($x$-axis) which together with $\n$ defines the
incidence plane, and the vector $\t$ ($y$-axis) looking toward the
reader, which is normal to the incidence plane.} \label{f2}
\end{figure}

Since reflection of mode 2 is in general nonspecular, it can
happen that the wave vectors of reflected and transmitted waves
will be arranged as shown in fig. \ref{f2}, and it follows that
there are two critical angles for $\varphi$. The first critical
angle $\varphi_{c1}$ ($q^2=1/\epsilon_0$) is the angle of total
reflection. At it the transmitted wave becomes evanescent. The
totally reflected field contains two modes. At the second critical
angle $\varphi_{c2}$, when $q$ is in the range
\begin{equation}\label{023}
1<q^2<\fr{(1+\eta)(1+\eta'cos^2(\theta_a))}{1+\eta'}.
\end{equation}
the reflected mode 1 also becomes evanescent. Together with
evanescent transmitted wave the mode 1 constitutes a surface wave,
propagating along the interface. In that case we have nonspecular
total reflection of the mode $\cE_2$.

\begin{figure}[tbh]
{\par\centering\resizebox*{8cm}{!}{\includegraphics{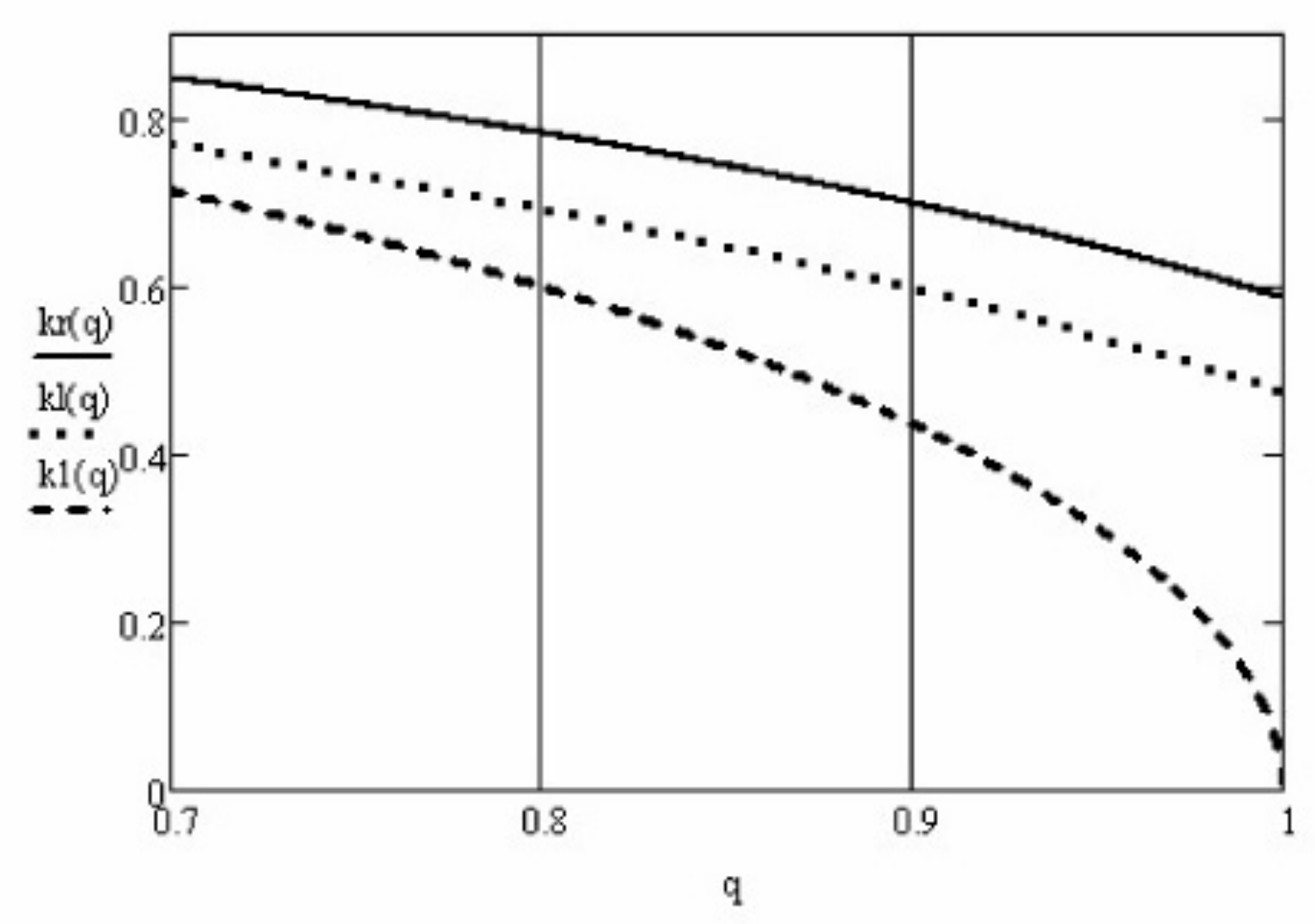}}\par}
\caption{Dependence of dimensionless normal components of incident
and reflected waves on $q=k\cos\varphi/k_0\sqrt{\epsilon_0}$. The
solid curve corresponds to the incident wave moving to the right
$kr(q)=k_{2r\bot}/k_0\sqrt{\epsilon_0}$. The dotted curve
corresponds to the reflected wave of mode 2 moving to the left
$kl(q)=k_{2l\bot}/k_0\sqrt{\epsilon_0}$. And the broken curve
corresponds to the reflected wave of mode 1 moving to the left
$k1(q)=k_{1l\bot}/k_0\sqrt{\epsilon_0}$. It is seen that at $q>1$
the mode 1 ceases to propagate. Its normal component
$k1(q)=\sqrt{1-q^2}=-i\sqrt{q^2-1}$ becomes imaginary, therefore
the reflected mode 1 becomes an evanescent wave. Together with
transmitted wave, which becomes evanescent at $q^2=1/\epsilon_0$,
the mode 1 constitute the surface electromagnetic wave.}
\label{f3}
\end{figure}

In figure \ref{f3} it is shown how do the normal components of
wave vectors change with increase of $q$, which is equivalent to
decrease of $\varphi$. For $\epsilon_0=1.6$ the first critical
angle corresponds to $q\approx0.8$. The second critical angle
corresponds to $q=1$.

\subsection{Reflection and refraction
from inside anisotropic medium}

The wave function in the full space is
\begin{equation}\label{23}
\Psi(\rr)=\Theta(z<0)\lt(e^{i\ora{\skk}_{j}\cdot\srr}\ora{\psi}_{j}+\sum\limits_{j'=1,2}
e^{i\ola{\skk}_{j'}\cdot\srr}\ola{\psi}_{j'}\ora\rho_{j'j}\rt)+
\Theta(z>0)e^{i\skk_0\cdot\srr}\Big(\psi_{e}\ora\tau_{ej}+\psi_{m}\ora\tau_{mj}\Big),
\end{equation}
where $\psi=\cE+\cH$, arrows show direction of waves propagation,
$\ora{\psi}_{j}$ denotes the incident wave of mode $j$ ($j=1,2$),
$\ola{\psi}_{j'}$ ($l=1,2$) denotes reflected wave of mode $j'$,
$\ora{\kk}_{j}=(\kk_\|,k_{jr\bot})$,
$\ola{\kk}_{j'}=(\kk_\|,-k_{j'l\bot})$,
$\kk_0=(\kk_\|,\sqrt{k_0^2-k_\|^2})$, $\psi_{e,m}$,
$\ora\tau_{e,mj}$ are fields and transmission amplitudes of TE-
and TM-modes for incident $j$-mode respectively. To find
reflection $\ora\rho$ and transmission $\ora\tau$ amplitudes (the
arrow over them shows the direction of propagation of the incident
wave toward the interface), we need to impose on \eref{23} the
following boundary conditions.

\subsection{General equations from boundary conditions}

Every incident wave field can be decomposed at the interface into
TE- and TM-modes. In TE-mode electric field is perpendicular to
the incidence plane, $\cE\propto\t$, therefore contribution of
$j$-th mode into TE-mode is $(\cE_j\cdot\t)$. In TM-mode magnetic
field is perpendicular to the incidence plane, $\cH\propto\t$,
therefore contribution of $j$-th mode into TM-mode is
$(\cH_j\cdot\t)$. For transmitted field in TE-mode we accept
$\ora\cE_e=\t$, $\ora\cH_e=[\ka_0\times\t]$, and for transmitted
field in TM-mode we accept $\ora\cE_m=\t$,
$\ora\cE_m=-[\ka_0\times\t]$.

\subsubsection{TE-boundary conditions}

In TE-mode for incident j-mode we have the following three
equations from boundary conditions:
\begin{enumerate}
    \item continuity of electric field
\begin{equation}\label{a23}
(\t\cdot\ora\cE_j)+(\t\cdot\ola\cE_1)\ora\rho_{1j}+(\t\cdot\ola\cE_2)\ora\rho_{2j}=\ora\tau_{ej},
\end{equation}
    \item continuity of magnetic field parallel to the interface
\begin{equation}\label{a23a}
(\lv\cdot\ora\cH_j)+(\lv\cdot\ola\cH_1)\ora\rho_{1j}+(\lv\cdot\ola\cH_2)\ora\rho_{2j}=(\lv\cdot[\ka_0\times\t])\ora\tau_{ej}\equiv
-\kappa_{0\bot}\ora\tau_{e1},
\end{equation}
    \item and continuity of the normal component of magnetic induction,
    which for $\mu=1$ looks
\begin{equation}\label{b23a}
(\n\cdot\ora\cH_j)+(\n\cdot\ola\cH_1)\ora\rho_{1j}+(\n\cdot\ola\cH_2)\ora\rho_{2j}=(\n\cdot[\ka_0\times\t])\ora\tau_{ej}\equiv
\kappa_{0\|}\tau_{ej}.
\end{equation}
\end{enumerate}
The last eq. \eref{b23a} is, in fact, not needed, because it
coincides with \eref{a23}. It is a good exercise to check the
identity of \eref{a23} and \eref{b23a} using explicit expressions
for $\cE$ and $\cH$. Some examples of such substitutions will be
presented in the Appendix A.

\subsubsection{TM-boundary conditions}

In TM-mode we have the equations
\begin{enumerate}
    \item continuity of magnetic field
\begin{equation}\label{c23}
(\t\cdot\ora\cH_j)+(\t\cdot\ola\cH_1)\ora\rho_{1j}+(\t\cdot\ola\cH_2)\ora\rho_{2j}=\ora\tau_{mj},
\end{equation}
\item continuity of electric field parallel to the interface
\begin{equation}\label{d23a}
(\lv\cdot\ora\cE_j)+(\lv\cdot\ola\cE_1)\ora\rho_{1j}+(\lv\cdot\ola\cE_2)\ora\rho_{2j}=-(\lv\cdot[\ka_0\times\t])\ora\tau_{mj}
\equiv\kappa_{0\bot}\ora\tau_{mj},
\end{equation}
\item and continuity of the normal component of field $\D$
    \begin{equation}\label{e23a}
(\n\cdot\eps\ora\cE_j)+(\n\cdot\eps\ola\cE_1)\ora\rho_{1j}+(\n\cdot\eps\ola\cE_2)\ora\rho_{2j}=
-(\n\cdot[\ka_0\times\t])\ora\tau_{mj}\equiv\kappa_{0\|}\ora\tau_{mj}.
\end{equation}
\end{enumerate}
Again we can neglect Eq. \eref{e23a}, because it coincides with
\eref{c23}, and again it is a good, though a little bit more
difficult, exercise to check identity of these equations using
explicit expressions for $\cE$ and $\cH$. In the following we will
not show third equations like {b23a} and \eref{e23a}, because they
are useless.

Exclusion of $\ora\tau_{ej}$ from \eref{a23} and \eref{a23a}, and
exclusion of $\ora\tau_{mj}$ from \eref{c23} and \eref{d23a} give
two equations for $\ora\rho_{1j}$, $\ora\rho_{2j}$, which is
convenient to represent in the matrix form {\small
\begin{equation}\label{m}
\left(%
\begin{array}{cc}
 \lt((\lv\cdot\ola\cH_1)+\kappa_{0\bot}(\t\cdot\ola\cE_1)\rt) &
 \lt((\lv\cdot\ola\cH_2)+\kappa_{0\bot}(\t\cdot\ola\cE_2)\rt) \\
\lt(\kappa_{0\bot}(\t\cdot\ola\cH_1)-(\lv\cdot\ola\cE_1)\rt) &
\lt(\kappa_{0\bot}(\t\cdot\ola\cH_2)-(\lv\cdot\ola\cE_2)\rt) \\
\end{array}%
\right){\ora\rho_{1j}\choose\ora\rho_{2j}}=-{(\lv\cdot\ora\cH_1)+\kappa_{0\bot}(\t\cdot\ora\cE_1)\choose
\kappa_{0\bot}(\t\cdot\ora\cH_j)-(\lv\cdot\ora\cE_j)}.
\end{equation}}
Solution of this equation is very simple if to take into account
that reciprocal of an arbitrary 2x2 matrix looks
\begin{equation}\label{m2}
\left(%
\begin{array}{cc}
  a & b \\
  c & d \\
\end{array}%
\right)^{-1}=\fr1{ad-bc}\left(%
\begin{array}{cc}
  d & -b \\
  -c & a \\
\end{array}%
\right).
\end{equation}
Therefore {\small
\begin{equation}\label{mm}
{\ora\rho_{1j}\choose\ora\rho_{2j}}=\fr1D\left(%
\begin{array}{cc}
 -\lt(\kappa_{0\bot}(\t\cdot\ola\cH_2)-(\lv\cdot\ola\cE_2)\rt)&
 \lt((\lv\cdot\ola\cH_2)+\kappa_{0\bot}(\t\cdot\ola\cE_2)\rt) \\
\lt(\kappa_{0\bot}(\t\cdot\ola\cH_1)-(\lv\cdot\ola\cE_1)\rt) &
-\lt((\lv\cdot\ola\cH_1)+\kappa_{0\bot}(\t\cdot\ola\cE_1)\rt)\\
\end{array}%
\right){(\lv\cdot\ora\cH_j)+\kappa_{0\bot}(\t\cdot\ora\cE_j)\choose
\kappa_{0\bot}(\t\cdot\ora\cH_j)-(\lv\cdot\ora\cE_j)},
\end{equation}}
Where $D$ is determinant
$$D=\lt((\lv\cdot\ola\cH_1)+\kappa_{0\bot}(\t\cdot\ola\cE_1)\rt)\lt(\kappa_{0\bot}
(\t\cdot\ola\cH_2)-(\lv\cdot\ola\cE_2)\rt)-$$
\begin{equation}\label{dt}
-\lt((\lv\cdot\ola\cH_2)+\kappa_{0\bot}(\t\cdot\ola\cE_2)\rt)\lt(\kappa_{0\bot}(\t\cdot\ola\cH_1)-(\lv\cdot\ola\cE_1)\rt).
\end{equation}
Substitution of these expressions into \eref{a23} and \eref{c23}
gives transmissions $\ora\tau_{e,mj}$
\begin{equation}\label{mm1}
{\ora\tau_{ej}\choose\ora\tau_{mj}}={(\t\cdot\ora\cE_j)\choose(\t\cdot\ora\cH_j)}+\left(%
\begin{array}{cc}
 (\t\cdot\ola\cE_1)&
 (\t\cdot\ola\cE_2) \\
(\t\cdot\ola\cH_1) &
(\t\cdot\ola\cH_2)\\
\end{array}%
\right){\ora\rho_{1j}\choose\ora\rho_{2j}},
\end{equation}

\subsubsection{The most general case}

Above we considered the case when the incident wave has
polarization vector $\e_{j}$ with unit amplitude. (We remind that
vectors $\e_{j}$ are not normalized to unity.) To find later
reflections from plain plates we will need a more general case,
when the incident wave has both modes with amplitudes $x_{1,2}$.
To find amplitudes of reflected and transmitted waves in the
general case it is convenient to represent the state of the
incident wave in the form of 2 dimensional vector
\begin{equation}\label{mx}
|\ora x\rangle={\ora x_1\choose \ora x_2}.
\end{equation}
then the states of reflected and transmitted waves are also
described by 2-dimensional vectors, which can be represented as
\begin{equation}\label{m10}
|\ola\psi\rangle={\ola \psi_1\choose \ola \psi_2}=\ora\Rho'|\ora
x\rangle,\qquad |\ora\psi_0\rangle={\ora \psi_e\choose \ora
\psi_m}=\ora\Ta'|\ora x\rangle,
\end{equation}
where $\ora\Rho'$ and $\ora\Ta'$ are two dimensional matrices
\begin{equation}\label{xa3a}
\ora\Rho'=\left(%
\begin{array}{cc}
  \ora\rho_{11} & \ora\rho_{12} \\
  \ora\rho_{21} & \ora\rho_{22} \\
\end{array}%
\right),\qquad \ora\Ta'=\left(%
\begin{array}{cc}
  \ora\tau_{e1} & \ora\tau_{e2} \\
  \ora\tau_{m1} & \ora\tau_{m2} \\
\end{array}%
\right).
\end{equation}
We introduced the prime here and below to distinguish transmission
and reflection from inside the medium from the similar matrices
obtained for incident waves outside the medium.

These formulas will be used later for calculation of reflection
and transmission of plain parallel anisotropic plates. In the case
of a plate we have two interfaces, therefore we need also
reflection and transmission at the left interface from inside and
outside the plate. Reflection and transmission from inside the
plate can be easily found from symmetry considerations. Their
representation is obtained from \eref{mm} --- \eref{mm1} by
reverse of arrows and change of the sign before $\kappa_{0\bot}$.
After this action we find
\begin{equation}\label{x1a3a}
\ola\Rho'=\left(%
\begin{array}{cc}
  \ola\rho_{11} & \ola\rho_{12} \\
  \ola\rho_{21} & \ola\rho_{22} \\
\end{array}%
\right),\qquad \ola\Ta'=\left(%
\begin{array}{cc}
  \ola\tau_{e1} & \ola\tau_{e2} \\
  \ola\tau_{m1} & \ola\tau_{m2} \\
\end{array}%
\right).
\end{equation}
Reflection from outside the medium is to be considered separately.

\subsubsection{Energy conservation}

It is always necessary to control correctness of the obtained
formulas. One of the best controls is the test of energy
conservation. One should always check whether the energy density
flux of incident wave along the normal to interface is equal to
the sum of energy density fluxes of reflected and transmitted
waves, and the most important in such tests is the correct
definition of the energy fluxes. In isotropic media it is possible
to define energy flux along a vector $\n$ as
\begin{equation}\label{jn}
(\J\cdot\n)=\fr{(\kk\cdot\n)}k\fr{c}{\sqrt{\epsilon}}\fr{\varepsilon
\cE^2+\cH^2}{8\pi},
\end{equation}
or
\begin{equation}\label{jn1}
(\J\cdot\n)=c\fr{(\n\cdot[\cE\times\cH])}{4\pi}.
\end{equation}
In isotropic media both definitions are equivalent, because
$\cH=[\kk\times\cE]$, and $(\kk\cdot\cE)=0$. The first definition
looks even more preferable since the second one can be written
even for stationary fields, where there are no energy flux.

In anisotropic media only the second definition is valid, and
because in mode 2 the field $\cE$ is not orthogonal to $\kk$, the
direction of the energy density flux is determined not only by
wave vector, but also by direction of the field $\cE$ itself.

\subsection{Reflection and refraction
from outside an anisotropic medium}

Let's consider the case, when the half space at $z<0$ is vacuum,
and that at $z>0$ is an anisotropic medium. The incident wave
falls onto interface from vacuum. The wave function in the full
space can be represented as
\begin{equation}\label{2a3}
\Psi(\rr)=\Theta(z<0)\Big(e^{i\ora\skk_{0}\srr}\ora\psi_j+e^{i\ola\skk_{0}\srr}\sum_{j'=e,m}\ola\psi_{j'}\ora\rho_{j'j}\Big)+
\Theta(z>0)\lt(e^{i\ora\skk_{1}\srr}\ora\psi_{1}\ora\tau_{1j}+e^{i\ora\skk_{2}\srr}\ora\psi_{2}\ora\tau_{2j}\rt),
\end{equation}
where $j,j'$ denote $e$ or $m$ for TE- and TM-modes respectively,
the term $\exp(i\ora\kk_{0}\rr)\ora\ps_{j}$ with the wave vector
$\ora\kk_{0}=(\kk_\|,k_{0\bot}=\sqrt{k_0^2-k_\|^2})$ describes the
plain wave incident on the interface from vacuum. In TE-mode
factor $\ora\ps_{e}=\ora\cE_{e}+\ora\cH_{e}$ contains
$\ora\cE_{e}=\t$ and $\ora\cH_{e}=[\ora\ka_0\t]$. In TM-mode
factor $\ora\ps_{m}=\ora\cE_{m}+\ora\cH_{m}$ contains
$\ora\cH_{e}=\t$ and $\ora\cE_{e}=-[\ora\ka_0\t]$.

The reflected wave has the wave vector
$\ola\kk_{0}=(\kk_\|,-k_{0\bot})$, and fields $\ola\cE_{e}=\t$,
$\ola\cH_{e}=[\ola\ka_0\t]$, $\ola\cH_{m}=\t$, and
$\ola\cE_{m}=-[\ola\ka_0\t]$. The refracted field contains two
wave modes with wave vectors $\ora\kk_{1}=(\kk_\|,k_{1\bot})$,
$\ora\kk_{2}=(\kk_\|,k_{2r\bot})$ and electric fields
$\ora\cE_{1}=\e_1=[\av\ora\ka_{1}]$ and
$\ora\cE_{2}=\e_2=\av-\ora\ka_{2}(\av\ora\ka_{2})\epsilon(\theta_{\ora2})/\epsilon_0$.
Here $\ka=\kk/k$, $k_{1\bot}=\sqrt{\epsilon_0k_0^2-k_\|^2}$, and
$k_{\ora2\bot}$ is given by \eref{19a}. For incident TE-mode
reflection $\rho_{ee}$, $\rho_{me}$ and refraction $\tau_{je}$
amplitudes ($j=1,2$) are found from boundary conditions
\begin{equation}\label{A23}
(\t\ora\cE_{1})\ora\tau_{1e}+(\t\ora\cE_{2})\ora\tau_{2e}=1+\ora\rho_{ee},
\end{equation}
\begin{equation}\label{A23a}
(\lv\ora\cH_{1})\ora\tau_{1e}+(\lv\ora\cH_{2})\ora\tau_{2e}=-\kappa_{0\bot}(1-\ora\rho_{ee}),
\end{equation}
\begin{equation}\label{C23}
(\t\ora\cH_{1})\ora\tau_{1e}+(\t\ora\cH_{2})\ora\tau_{2e}=\ora\rho_{me},
\end{equation}
\begin{equation}\label{d223a}
(\lv\ora\cE_{1})\ora\tau_{1e}+(\lv\ora\cE_{2})\ora\tau_{2e}=-\kappa_{0\bot}\ora\rho_{me}.
\end{equation}
Exclusion of $\ora\rho_{ee}$ and $\ora\rho_{me}$ leads to
\begin{equation}\label{ee}
\left(%
\begin{array}{cc}
 \lt(\kappa_{0\bot}(\t\ora\cE_{1})-(\lv\ora\cH_{1})\rt) & \lt(\kappa_{0\bot}(\t\ora\cE_{2})-
(\lv\ora\cH_{2})\rt) \\
 \lt(\kappa_{0\bot}(\t\ora\cH_{1})+(\lv\ora\cE_{1})\rt)  & \lt(\kappa_{0\bot}(\t\ora\cH_{2})+(\lv\ora\cE_{2})\rt) \\
\end{array}%
\right){\ora\tau_{1e}\choose
\ora\tau_{2e}}={2\kappa_{0\bot}\choose 0}
\end{equation}
and the solution
\begin{equation}\label{ee1}
{\ora\tau_{1e}\choose
\ora\tau_{2e}}=\fr1{D_e}\left(\begin{array}{cc}
 \lt(\kappa_{0\bot}(\t\ora\cH_{2})+(\lv\ora\cE_{2})\rt) & -\lt(\kappa_{0\bot}(\t\ora\cE_{2})-
(\lv\ora\cH_{2})\rt) \\
 -\lt(\kappa_{0\bot}(\t\ora\cH_{1})+(\lv\ora\cE_{1})\rt)  &\lt(\kappa_{0\bot}(\t\ora\cE_{1})-(\lv\ora\cH_{1})\rt)
  \\
\end{array}%
\right){2\kappa_{0\bot}\choose 0}
\end{equation}
where $D_e$ is determinant
$$D_e=\lt(\kappa_{0\bot}(\t\ora\cE_{1})-(\lv\ora\cH_{1})\rt)\lt(\kappa_{0\bot}(\t\ora\cH_{2})+(\lv\ora\cE_{2})\rt)-$$
\begin{equation}\label{ee2}
-\lt(\kappa_{0\bot}(\t\ora\cE_{2})-
(\lv\ora\cH_{2})\rt)\lt(\kappa_{0\bot}(\t\ora\cH_{1})+(\lv\ora\cE_{1})\rt).
\end{equation}
Substitution of $\ora\tau_{je}$ into \eref{A23} and \eref{C23}
gives
\begin{equation}\label{ee3}
{\ora\rho_{ee}\choose\ora\rho_{me}}=
\left(%
\begin{array}{cc}
  (\t\ora\cE_{1}) & (\t\ora\cE_{2}) \\
  (\t\ora\cH_{1}) & (\t\ora\cH_{2}) \\
\end{array}%
\right){\ora\tau_{1e}\choose \ora\tau_{2e}}-{1\choose0}.
\end{equation}

In the case of incident TM-mode we have boundary conditions
\begin{equation}\label{A23aa}
(\t\ora\cH_{1})\ora\tau_{1m}+(\t\ora\cH_{2})\ora\tau_{2m}=1+\ora\rho_{mm},
\end{equation}
\begin{equation}\label{A2a3a}
(\lv\ora\cE_{1})\ora\tau_{1m}+(\lv\ora\cE_{2})\ora\tau_{2m}=\kappa_{0\bot}(1-\ora\rho_{mm}),
\end{equation}
\begin{equation}\label{C23a}
(\t\ora\cE_{1})\ora\tau_{1m}+(\t\ora\cE_{2})\ora\tau_{2m}=\ora\rho_{em},
\end{equation}
\begin{equation}\label{d2a23a}
(\lv\ora\cH_{1})\ora\tau_{1m}+(\lv\ora\cH_{2})\ora\tau_{2m}=\kappa_{0\bot}\ora\rho_{em}.
\end{equation}
Exclusion of $\ora\rho_{me}$ and $\ora\rho_{mm}$ leads to
\begin{equation}\label{me}
\left(%
\begin{array}{cc}
 \lt(\kappa_{0\bot}(\t\ora\cH_{1})+(\lv\ora\cE_{1})\rt) &
 \lt(\kappa_{0\bot}(\t\ora\cH_{2})+
(\lv\ora\cE_{2})\rt) \\
 \lt(\kappa_{0\bot}(\t\ora\cE_{1})-(\lv\ora\cH_{1})\rt)  & \lt(\kappa_{0\bot}(\t\ora\cE_{2})-(\lv\ora\cH_{2})\rt) \\
\end{array}%
\right){\ora\tau_{1m}\choose
\ora\tau_{2m}}={2\kappa_{0\bot}\choose 0}.
\end{equation}
Therefore
\begin{equation}\label{me1}
{\ora\tau_{1m}\choose
\ora\tau_{2m}}=\fr1{D_m}\left(\begin{array}{cc}
 \lt(\kappa_{0\bot}(\t\ora\cE_{2})-(\lv\ora\cH_{2})\rt) &
 -\lt(\kappa_{0\bot}(\t\ora\cH_{2})+
(\lv\ora\cE_{2})\rt) \\
 -\lt(\kappa_{0\bot}(\t\ora\cE_{1})-(\lv\ora\cH_{1})\rt)  &\lt(\kappa_{0\bot}(\t\ora\cH_{1})+(\lv\ora\cE_{1})\rt)
  \\
\end{array}%
\right){2\kappa_{0\bot}\choose 0},
\end{equation}
where $D_m=-D_e$ \eref{ee2}. Substitution of $\ora\tau_{jm}$ into
\eref{A23aa} and \eref{C23a} gives
\begin{equation}\label{me2}
{\ora\rho_{em}\choose\ora\rho_{mm}}=\left(%
\begin{array}{cc}
  (\t\ora\cE_{1})& (\t\ora\cE_{2}) \\
  (\t\ora\cH_{1}) & (\t\ora\cH_{2})\\
\end{array}%
\right){\ora\tau_{1m}\choose\ora\tau_{2m}}-{0\choose1}.
\end{equation}
In the general case, when the incident wave has an amplitude
$\ora\xi_e$ in TE-mode and amplitude $\ora\xi_m$ in TM-mode, the
state of the incident wave can be described by two-dimensional
vector
\begin{equation}\label{me3}
|\ora\xi_0\rangle={\ora\xi_e\choose\ora\xi_m},
\end{equation}
and the states of reflected and transmitted waves can be
represented as
\begin{equation}\label{me4}
|\ola\xi_0\rangle={\ola\xi_e\choose \ola\xi_m}=\ora\Rho|\ora
\xi_0\rangle,\qquad |\ora\xi\rangle={\ora \xi_1\choose \ora
\xi_2}=\ora\Ta|\ora\xi_0\rangle,
\end{equation}
where $\ora\Rho$ and $\ora\Ta$ are the two dimensional matrices
\begin{equation}\label{xa3aa}
\ora\Rho=\left(%
\begin{array}{cc}
  \ora\rho_{ee} & \ora\rho_{em} \\
  \ora\rho_{me} & \ora\rho_{mm} \\
\end{array}%
\right),\qquad \ora\Ta=\left(%
\begin{array}{cc}
  \ora\tau_{1e} & \ora\tau_{1m} \\
  \ora\tau_{2e} & \ora\tau_{2m} \\
\end{array}%
\right).
\end{equation}

\section{Reflection and transmission of a plain parallel plate of thickness $L$}

Now, when we understand what happens at interfaces, we can
construct~\cite{igu} expressions for reflection, $\ora\hR(L)$, and
transmission, $\ora\hT(L)$, matrices for a whole anisotropic plain
parallel plate of some thickness $L$, when the state of the
incident wave is described by a general vector $\ora\xi_0\rangle$.
To do that let's denote the state of field of the modes $\e_1$ and
$\e_2$ incident from inside the plate onto the second interface at
$z=L$ by unknown  2-dimensional vector $|\ora x\rangle$ \eref{mx}.
If we were able to find $|x\rangle$ we could immediately write the
state of transmitted field
\begin{equation}\label{ax}
\ora\hT(L)|\ora\xi_0\rangle=\ora\Ta'|\ora x\rangle,
\end{equation}
and the state of the field, reflected from the whole plate
\begin{equation}\label{ksx}
\ora\hR(L)|\ora\xi_0\rangle=\ora\Rho|\xi_0\rangle+\ola\Ta'\ola\hE(L)\ora\Rho'|\ora
x\rangle,
\end{equation}
where $\ola\hE(L)$, $\ora\hE(L)$ denote diagonal matrices
\begin{equation}\label{2x}
\ola\hE(L)=\left(%
\begin{array}{cc}
  \exp(i k_{1\bot}L) & 0 \\
 0 & \exp(ik_{2l\bot}L)\\
\end{array}%
\right),\qquad \ola\hE(L)=\left(%
\begin{array}{cc}
  \exp(ik_{1\bot}L) & 0 \\
 0 & \exp(ik_{2r\bot}L)\\
\end{array}%
\right).
\end{equation}
which describe propagation of two modes between two interfaces.
Here $k_{1\bot}=\sqrt{\epsilon_0k_0^2-k_\|^2}$, while $
k_{2r\bot}$ and $k_{2l\bot}$ are calculated according to \eref{19}
or \eref{19a} and \eref{21}, respectively.
\begin{figure}[t!]
{\par\centering\resizebox*{8cm}{!}{\includegraphics{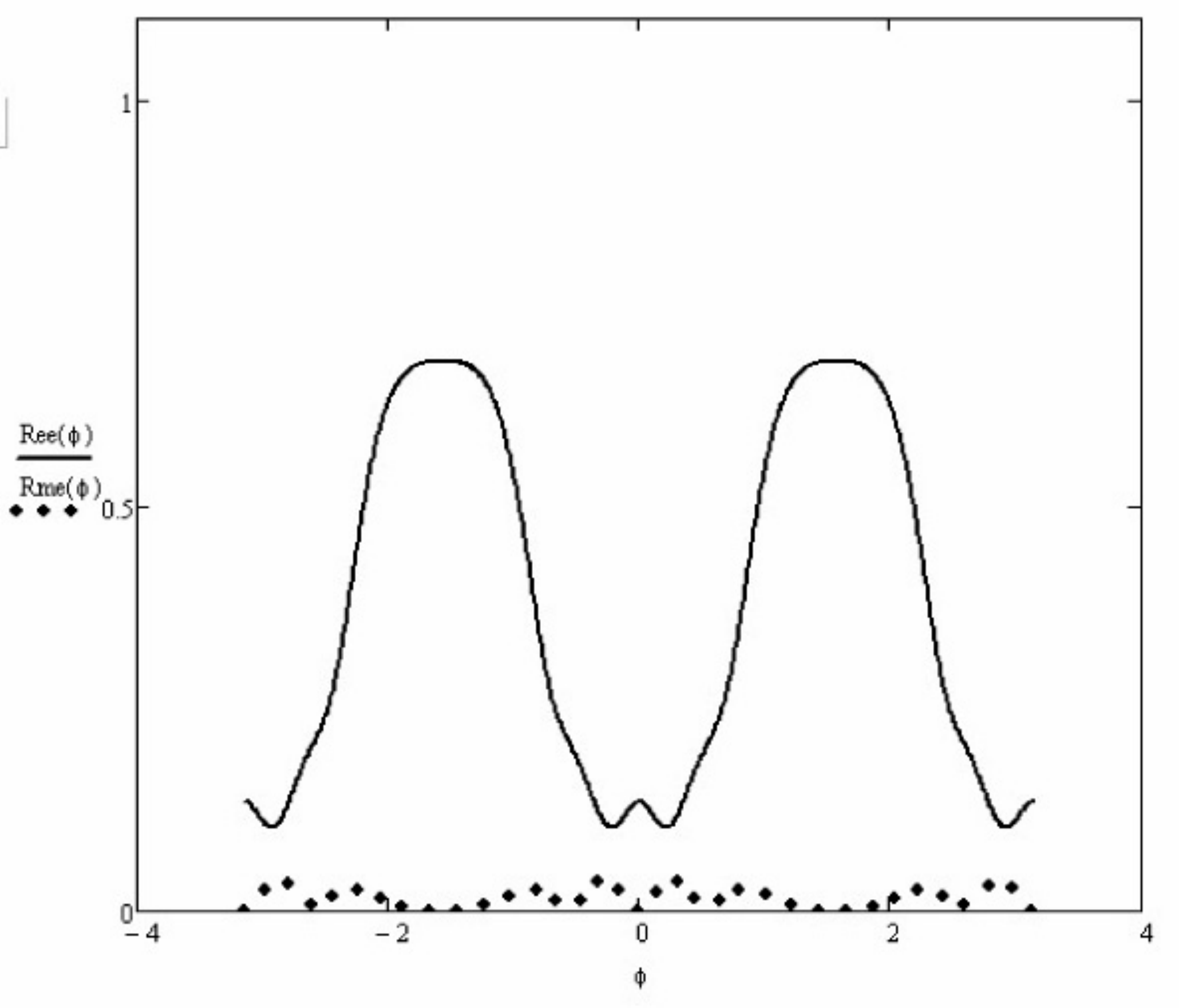}}\par}
\caption{Dependence of reflectivities $|R_{ee}|^2$ and
$|R_{me}|^2$ of an anisotropic plate with $\epsilon_0=1.6$,
$\eta=0.8$ and dimensionless thickness $L\omega/c=10$ on angle
$\phi$ of the plate rotation around its normal, when the
anisotropy vector $\av$ is parallel to interfaces and at $\phi=0$
is directed along $\kk_\|$. The incidence angle $\theta$ is given
by $\sin\theta=0.9$.} \label{r}
\end{figure}

It is very easy to put down a self consistent equation for
determination of $|\ora x\rangle$:
\begin{equation}\label{ksx2}
|\ora x\rangle=\ora\hE(L)\ora\Ta|\xi_0\rangle+\ora\hE(L)
\ola\Rho'\ola\hE(L)\ora\Rho'|\ora x\rangle.
\end{equation}
The first term at the right hand side describes the incident state
transmitted through the first interface and propagated up to the
second one. The second term describes contribution to the state
$|\ora x\rangle$ of the $|\ora x\rangle$ itself. After reflection
from the second interface this state propagates to the left up to
the first interface, and after reflection from it propagates back
to the point $z=L$. Two terms at the right hand side of \eref{ksx}
add together, which results to some new state. But we denoted it
$|\ora x\rangle$, and it explains derivation of the equation
\eref{ksx}.

From \eref{ksx} we can directly find
\begin{equation}\label{ksx1}
|\ora x\rangle=\Big[\hI-\ora\hE(L)
\ola\Rho'\ola\hE(L)\ora\Rho'\Big]^{-1}\ora\hE(L)\ora\Ta|\xi_0\rangle,
\end{equation}
and substitution into \eref{ax} and \eref{ksx} gives
\begin{equation}\label{kx2}
\ora\hT(L)\equiv\left(%
\begin{array}{cc}
  T_{ee} & T_{em} \\
  T_{me} & T_{mm} \\
\end{array}%
\right)=\ora\Ta'\Big[\hI-\ora\hE(L)
\ola\Rho'\ola\hE(L)\ora\Rho'\Big]^{-1}\ora\hE(L)\ora\Ta,
\end{equation}
\begin{equation}\label{kx3}
\ora\hR(L)\equiv\left(%
\begin{array}{cc}
  R_{ee} & R_{em} \\
  R_{me} & R_{mm} \\
\end{array}%
\right)=\ora\Rho+\ola\Ta'\ola\hE(L)\ora\Rho'\Big[\hI-\ora\hE(L)
\ola\Rho'\ola\hE(L)\ora\Rho'\Big]^{-1}\ora\hE(L)\ora\Ta.
\end{equation}

\begin{figure}[t!]
{\par\centering\resizebox*{8cm}{!}{\includegraphics{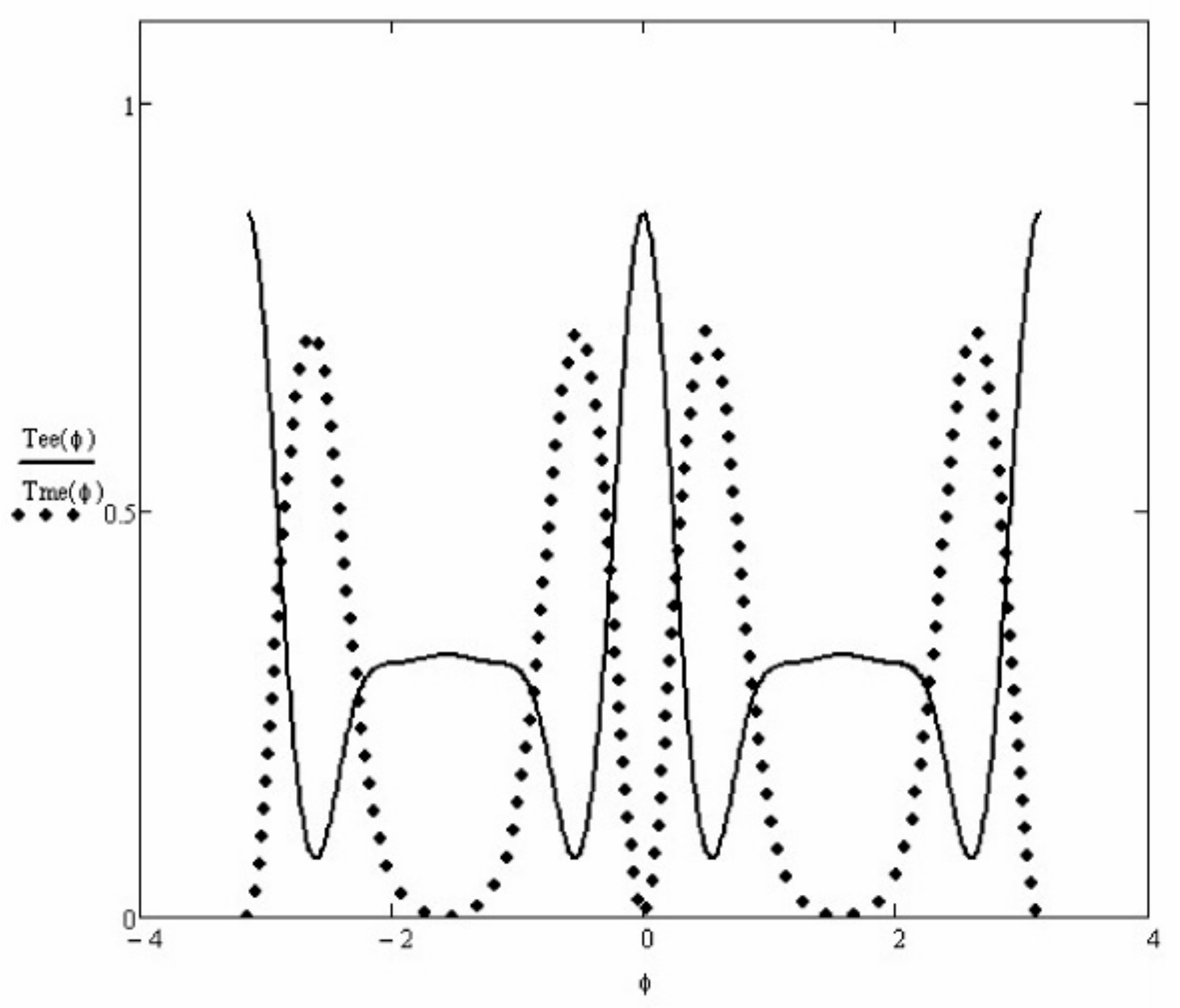}}\par}
\caption{Dependence of transmissivities $|T_{ee}|^2$ and
$|T_{me}|^2$ of an anisotropic plate on angle $\phi$ of the plate
rotation around its normal, with all the parameters the same as
shown in caption of fig.\ref{r}.} \label{t}
\end{figure}

With these formulas we can easily calculate all the reflectivities
and transmissivities for arbitrary parameters, arbitrary incidence
angles, arbitrary incident polarizations and arbitrary direction
of the anisotropy vector $\av$. In fig.\ref{r} we present, for
example, reflectivities of TE-mode wave from a plate of thickness
$L$ such, that $L\omega/c=10$. The anisotropy vector is parallel
to interfaces. Therefore, its orientation with respect to wave
vector $\kk_0$ of the incident wave varies with rotation of the
plate by an angle $\phi$ around its normal. The transmissivities
of the same plate in dependence on the angle $\phi$ are presented
in fig.\ref{t}.

It is important to note that after transmission through
anisotropic plate the transmitted ray has the same angle with the
plate normal as the incident one, and it is interesting to
investigate how the image of a real object splits, when observed
through a birefringent ideal plain plate.

\section{Conclusion}

We have shown how to find polarizations of plain waves propagating
in a single axis anisotropic media, how to calculate reflection
and refraction at an interface between anisotropic and isotropic
media, and how to calculate reflection and transmission of plain
parallel transparent anisotropic layers. We considered media
without absorption, but inclusion of absorption does not provide
any problem.

We have shown that reflection from interfaces in anisotropic media
is accompanied beam splitting, that reflection of the mixed mode
is nonspecular and can be characterized by two critical angles.
The first critical angle $\varphi_{c1}$ corresponds to total
reflection with nonspecular double beam splitting. The second
critical angle corresponds to total nonspecular reflection of
mixed mode without beam splitting and creation of surface
electromagnetic wave. The effect can be observed with the help of
birefringent cone, when the incident beam is transmitted through
the side surface and after reflection from the base plane goes out
of the cone.

We considered only one anisotropy vector, and it will be a good
exercise to consider whether addition of a second anisotropy axis
will not bring some new effects.

\appendix

\section{Substitution of mode polarizations in boundary conditions for incident wave of
mode 2}

\subsection{TE-mode}
Substitution of \eref{1c2} for $\cE_2$, $[\av\times\ka]$ for
$\cE_1$ and \eref{16a} for $\cH_{1,2}$ into \eref{a23} and
\eref{a23a} gives
$$(\t\cdot\av)+(\t\cdot[\av\times\ola\ka_1])\ora\rho_{12}+(\t\cdot\av)\ora\rho_{22}=
\ora\tau_{e2},\eqno(\ref{a23}a)$$
$$k_{2r\bot}(\t\cdot\av)-
k_{1\bot}(\t\cdot[\av\times\ola\ka_1])\ora\rho_{12}-k_{2l\bot}(\t\cdot\av)\ora\rho_{22}=k_{0\bot}\ora\tau_{e2},
\eqno(\ref{a23a}a)$$ where $\ola\ka_1=(\kk_\|,-k_{1\bot})/k_1$,
$k_1=k_0\sqrt{\epsilon_0}$, $k_{1\bot}=\sqrt{k_1^2-k_\|^2}$, and
$k_{2r,l\bot}$ are given in \eref{19a}, \eref{21}. In equation
(\ref{a23}a) we took into account that vector $\t$ is orthogonal
to $\ka$, and in (\ref{a23a}a) we used representation:
$\ka=\kappa_\|\lv+\kappa_\bot\n$, and the relation
$[\lv\times\n]=-\t$, which for any vector $\Av$ gives
\begin{equation}\label{c}
(\lv\cdot[\ka\times\Av])=([\lv\times\ka]\cdot\Av)=-\kappa_\bot(\t\cdot\Av).
\end{equation}
After exclusion of $\tau_{e2}$ from these equations we obtain the
equation
\begin{equation}\label{f2b3}
(\t\cdot[\av\times\ola\ka_1])(k_{1\bot}+k_{0\bot})\ora\rho_{12}+(\t\cdot\av)(k_{0\bot}+k_{2l\bot})\ora\rho_{22}
=(\t\cdot\av)(k_{2r\bot}-k_{0\bot}).
\end{equation}
\paragraph{In the case of isotropic medium} we have $k_{2r\bot}=k_{2l\bot}=k_{1\bot}=k_{\bot}$,
$\ola\ka_1=\ka_{l}=(\kk_\|,-k_\bot)/k$, therefore \eref{f2b3} is
reduced to
\begin{equation}\label{f2a3}
\fr{(\t\cdot[\av\times\ka_{r}])}{(\t\cdot\av)}\ora\rho_{12}+\ora\rho_{22}=\rho_{e0},
\qquad\ora\tau_{e2}=(\t\cdot\av)\tau_{e0},
\end{equation}
where
\begin{equation}\label{f2c3}
\rho_{e0}=\fr{(k_{\bot}-k_{0\bot})}{(k_{\bot}+k_{0\bot})},\qquad\tau_{e0}=1+\rho_{e0}=
\fr{2k_{\bot}}{(k_{\bot}+k_{0\bot})}
\end{equation}
are the standard reflection and transmission
amplitudes~\cite{land} of a pure TE-mode at an interface between
isotropic media.

{\bf If $\av=\t$} we have the typical incident TE-mode, and
$\rho_{12}$ is excluded because $(\t\cdot[\ka_{r}\times\av])=0$.

{\bf If $(\av\cdot\t)=0$} we have the typical incident TM-mode,
and from (\ref{a23}a), (\ref{a23a}a) it follows that $\rho_{12}=0$
\subsection{TM-mode}
The similar substitutions of $\cH_{1,2}$ into \eref{c23} for
TM-mode gives
$$\fr{k_{2r}}{k_0}(\t\cdot[\ora\ka_{2}\times\av])+
\fr{k_{1}}{k_0}(\t\cdot\av)\ora\rho_{12}+\fr{k_{2l}}{k_0}(\t\cdot[\ola\ka_{2}\times\av])\ora\rho_{22}=\ora\tau_m.
\eqno(\ref{c23}a)$$

For substitution of $\cE$ into \eref{d23a} we represent \eref{1c2}
in the form
\begin{equation}\label{g23}
\e_2=[\ka\times[\av\times\ka]]\fr{\epsilon_1(\theta)}{\epsilon_0}-\av\fr{\Delta\epsilon(\theta)}{\epsilon_0},
\end{equation}
where $\Delta\epsilon(\theta)=\epsilon_1(\theta)-\epsilon_0$, and
with account of \eref{c} we obtain
$$
\kappa_{2r\bot}(\t\cdot[\ora\ka_{2}\times\av])\fr{\epsilon_1(\theta_{2r})}{\epsilon_0}-
\fr{\Delta\epsilon(\theta_{2r})}{\epsilon_0}(\lv\cdot\av)-\kappa_{1\bot}(\t\cdot\av)\ora\rho_{12}-
$$
$$-\lt(
\kappa_{2l\bot}(\t\cdot[\ola\ka_{2}\times\av])\fr{\epsilon_1(\theta_{2l})}{\epsilon_0}+
\fr{\Delta\epsilon(\theta_{2l})}{\epsilon_0}(\lv\cdot\av)\rt)\ora\rho_{22}=
\kappa_{0\bot}\ora\tau_m. \eqno(\ref{d23a}a)$$

Exclusion of $\ora\tau_m$ with the help of (\ref{c23}b) gives
$$(\epsilon_0k_{0\bot}k_{2r}+k_{1\bot}k_{1})
(\t\cdot\av)\ora\rho_{12}+\Big(
(\epsilon_0k_{0\bot}k_{2r}+k_{2l\bot}k_{2l})
(\t\cdot[\ola\ka_{2}\times\av]) +
k_0^2\Delta\epsilon(\theta_{2l})(\lv\cdot\av)\Big)\ora\rho_{22}=$$
\begin{equation}\label{d2a3}
=k_{2r}(k_{2r\bot}-\epsilon_0k_{0\bot})(\t\cdot[\ora\ka_{2}\times\av])-
k_0^2\Delta\epsilon(\theta_{2r})(\lv\cdot\av).
\end{equation}
Together with \eref{f2b3} we have two equations for determination
of $\rho_{22}$ and $\rho_{12}$.
\paragraph{In the case of isotropic medium} we have $k_{2r\bot}=k_{2l\bot}=k_{1\bot}=k_{\bot}$,
$k_{2r}=k_{2l}=k_{1l}=k$, $\Delta\epsilon=0$,
$\epsilon_1=\epsilon_0$, $\ora\ka_{2}=\ora\ka=(\kk_\|,k_\bot)/k$,
$\ola\ka_{2}=\ola\ka_1=\ola\ka=(\kk_\|,-k_\bot)/k$. Therefore
\eref{d2a3} is reduced to
\begin{equation}\label{d2a3a}
(\t\cdot\av)\ora\rho_{12}+(\t\cdot[\ola\ka\times\av])\ora
\rho_{22}=(\t\cdot[\ora\ka\times\av])
\fr{(k_{\bot}-\epsilon_0k_{0\bot})}{(k_{\bot}+\epsilon_0k_{0\bot})}\equiv(\t\cdot[\ora\ka\times\av])\rho_{m0},
\end{equation}
and solution of \eref{d2a3a} with \eref{f2a3} gives
\begin{equation}\label{d2b3}
\ora\rho_{12}=
\fr{(\t\cdot\av)\lt((\t\cdot[\ora\ka\times\av])\rho_{m0}-
(\t\cdot[\ola\ka\times\av])\rho_{e0}\rt)}{(\t\cdot[\ola\ka\times\av])^2
+(\t\cdot\av)^2},
\end{equation}
\begin{equation}\label{d2c3}
\ora\rho_{22}=
\fr{(\t\cdot[\ola\ka\times\av])(\t\cdot[\ora\ka\times\av])\rho_{m0}+(\t\cdot\av)^2\rho_{e0}}{(\t\cdot[\ola\ka\times\av])^2
+(\t\cdot\av)^2}.
\end{equation}
For $\tau_m$ it follows from (\ref{c23}a) that
$$
\ora\tau_m=(\t\cdot[\ora\ka\times\av])(1+\rho_{m0})=(\t\cdot[\ora\ka\times\av])\fr{2k_{\bot}}{k_{\bot}+\epsilon_0k_{0\bot}}\equiv
(\t\cdot[\ora\ka\times\av])\tau_{m0}, \eqno(\ref{d23a}b)$$ where
$\rho_{m0}$ and $\tau_{m0}$ are the standard reflection and
transmission amplitudes~\cite{land} of a pure TM-mode at an
interface between isotropic media.

Thus, for an internal reflection from an interface of isotropic
dielectric with vacuum we obtained reflection and refraction
amplitudes for an incident field with an arbitrary polarization
$\e=[\ka\times[\av\times\ka]]$, which is determined by some unit
vector $\av$. If $\av=\t$, then $\ora\rho_{12}=0$ and
$\ora\rho_{22}=\rho_{e0}$. If $(\av\cdot\t)=0$, then again
$\ora\rho_{12}=0$, but $\ora\rho_{22}=
\rho_{m0}(\t\cdot[\ora\ka\times\av])/(\t\cdot[\ola\ka\times\av])$.
If $\av=\ola\ka$, then $\ora\rho_{22}$ is divergent, however the
reflected field $\ola\cE_2\ora\rho_{22}$ has the finite value
$\rho_{m0}(\t\cdot[\ora\ka\times\av])$.

\section{History of submission and rejection}

The paper was submitted to Am.J.Phys on September 1 of 2010. It
was rejected on September 24 because of negative reports of two
referees. The first referee said that he is lazy to read the
manuscript with pencil, but he saw that sections 2 and 4 are
absolutely not needed, because everything about reflections is
much better said in textbooks by Jackson and Griffith. The second
referee rejected because we, he said, erroneously told that the
recent paper was publish long ago at 1977. He said that since then
there were many papers on optical reflection and transmission.

If we could reply to referee we would mention that in textbooks by
Jackson and Griffith there are no word on anisotropic media. The
referee overlooked the main point of our paper. As for claim of
the second referee, we would like to say, that he can try to seek
on AJP home page a paper with key words ``electromagnetic waves in
anisotropic media.'' Then he will find the first article published
in 1977. So the second referee also overlooked the main point of
our article.


\begin{thebibliography}{99}
\bibitem[*]{mail}
e-mail: v.ignatovi@gmail.com
\bibitem{fed}F.I.Fedorov, {\it Optics of anisotropic media}, Minsk, BSSR
Ac.Sc., 1958. Eq. (20.4)
\bibitem{kuz}
Petr Ku\u{z}hel, ``Lecture 8: Light propagation in anisotropic
media'',\\ http://www.fzu.cz/~kuzelp/Optics/Lectures.htm;\\
http://www.fzu.cz/~kuzelp/Optics/Lecture8.pdf
\bibitem{kuz1}
Petr Ku\u{z}hel, {\it Electromagnetisme des milieux continus.
``Optique''}, Universite Paris-Nord, 2000/2001.
\bibitem{dit}R.W.Ditchburn. {\it Light}, Dover Publications Inc.N.Y. 1991.
\bibitem{land}L. D. Landau, E. M. Lifshitz and L. P. Pitaevskii,
Electrodynamics of Continuous Media. Second Edition: Volume 8
(Course of Theoretical Physics) Elsevier Butterworth-Heinemann,
2004. Ch.XI.
\bibitem{kunz} K.S.Kunz. ``Treatment of optical propagation in crystals using projection
dyadics.'' Am.J.Phys. {\bf 45} 267 (1977).
\bibitem{igl}
Vladimir K. Ignatovich, Loan T. N. Phan, Those wonderful elastic
waves. Am. J. Phys. {\bf 77} 1162 (2009)
\bibitem{igu}
Vladimir K. Ignatovich, Masahiko Utsuro, {\it Handbook on Neutron
Optics}, Wiley-VCN Verlag GmbH, \& Co. KGaA, 2009.
\end{thebibliography}
\end{document}